\documentclass[aps,pre,preprint,superscriptaddress]{revtex4-2} %

\makeatletter

\newif\ifuserevtex
\newif\ifusetwocolfigs

\@ifclassloaded{revtex4-2}{%
  \userevtextrue
  \@ifclasswith{revtex4-2}{reprint}{%
    \usetwocolfigstrue
  }{%
    \usetwocolfigsfalse
  }%
}{%
  \userevtexfalse
  \usetwocolfigsfalse %
}

\makeatother

\newif\ifshowlinenumbers
\usepackage[T1]{fontenc}         %
\usepackage[utf8]{inputenc}      %
\usepackage[english]{babel}      %

\usepackage{amssymb}             %
\usepackage{mathtools}           %
\usepackage{amsmath}%
\usepackage{commath}
\usepackage{csquotes}
\usepackage{mathrsfs}            %

\usepackage{dcolumn}             %

\usepackage{graphicx}            %
\graphicspath{{./}}

\usepackage{color,xcolor}        %
\usepackage{soul}                %
\usepackage{cancel}              %
\usepackage{comment}             %
\usepackage{moreverb}            %
\usepackage[normalem]{ulem}      %

\ifuserevtex

\else
    \usepackage{amsthm}
    
    \usepackage[a4paper,margin=2.3cm]{geometry} %
    \usepackage{appendix} %
    
    \usepackage[numbers]{natbib} %
    \usepackage{authblk} 
    \usepackage{setspace} %
\fi

\usepackage{pifont}

\usepackage{etoolbox}  %
\usepackage{environ}   %
\usepackage{xparse}
\usepackage{scalerel}

\usepackage[
  colorlinks=true,
  linkcolor=blue,  %
  citecolor=blue,  %
  urlcolor=blue    %
]{hyperref}

\usepackage[capitalise]{cleveref}  %

\usepackage{graphicx} \showlinenumbersfalse  %

\newif\ifshowallemails
\showallemailstrue  %

\makeatletter

\def\affillist{}
\newcommand{\myaffil}[2]{%
  \expandafter\def\csname affil@#1\endcsname{#2}%
  \ifuserevtex\else
    \listgadd{\affillist}{#1}%
  \fi
}

\newcommand{\emitoneaffil}[1]{%
  \expandafter\affil\expandafter[#1]{\csname affil@#1\endcsname}%
}

\newcommand{\orcidlink}[1]{%
  \href{https://orcid.org/#1}{\textsuperscript{\textcolor{green!50!black}{\textbullet}}}%
}

\NewDocumentCommand{\myauthor}{O{} O{} O{} m m}{%
  \ifuserevtex
    \author{#4%
      \ifx&#2&\else\ \orcidlink{#2}\fi
    }%
    \ifx&#1&\else
      \ifshowallemails
        \email{#1}%
      \else
        \ifx&#3&\else  %
          \email{#1}%
        \fi
      \fi
    \fi
    \@for\aid:=#5\do{%
      \expandafter\affiliation\expandafter{\csname affil@\aid\endcsname}%
    }%
  \else
    \author[#5]{#4%
      \ifx&#2&\else\ \orcidlink{#2}\fi
      \ifx&#1&\else
        \ifx&#3&%
          \ifshowallemails
            \thanks{\href{mailto:#1}{#1}}%
          \fi
        \else
          \thanks{Corresponding author: \href{mailto:#1}{#1}}%
        \fi
      \fi
    }%
  \fi
}

\newcommand{\storedabstract}{}  %
\newcommand{\storedkeywords}{}  %

\newcommand{\myabstract}[1]{%
  \long\def\storedabstract{#1}%
}

\newcommand{\mykeywords}[1]{%
  \def\storedkeywords{#1}%
}

\newcommand{\maketitleandabstract}{%
  \ifuserevtex
    \begin{abstract}
    \storedabstract
    \end{abstract}
    \keywords{\storedkeywords}
    \maketitle
  \else
    \maketitle
    \section*{Abstract}
    \storedabstract
    \ifx\storedkeywords\@empty
    \else
      \vspace{1em}
      \par\noindent \textbf{Keywords:} \storedkeywords
    \fi
  \fi
}

\newcommand{\storedacknowledgements}{}  %

\newcommand{\insertacknowledgements}{%
  \ifx\storedacknowledgements\@empty
  \else
    \ifuserevtex
      \begin{acknowledgments}
      \storedacknowledgements
      \end{acknowledgments}
    \else
      \section*{Acknowledgements}
      \storedacknowledgements
    \fi
  \fi
}

\makeatother

\ifshowlinenumbers
  \usepackage{lineno}
  \AtBeginDocument{%
    \ifuserevtex
      \ifusetwocolfigs
      \else
        \linenumbers
      \fi
    \else
      \linenumbers
    \fi
  }
\fi

\makeatletter

\def\onecolfig{\@ifnextchar[{\onecolfig@opt}{\onecolfig@opt[]}}
\def\endonecolfig{\end{figure}}
\def\onecolfig@opt[#1]{\begin{figure}[#1]}

\def\twocolfig{\@ifnextchar[{\twocolfig@opt}{\twocolfig@opt[]}}
\def\endtwocolfig{\ifusetwocolfigs \end{figure*} \else \end{figure} \fi}
\def\twocolfig@opt[#1]{%
  \ifusetwocolfigs
    \begin{figure*}[#1]%
  \else
    \begin{figure}[#1]%
  \fi
}

\makeatother

\usepackage{xr}

\makeatletter
\newcommand*{\addFileDependency}[1]{%
  \typeout{(#1)}
  \@addtofilelist{#1}
  \IfFileExists{#1}{}{\typeout{No file #1.}}
}
\makeatother

\newcommand{\papertitle}{The fracture resistance of elastic networks increases with the density of defects like a random walk}

\newcommand{\ie}{{\textit{i.e.}}}

\newcommand{\fraction}{{\nu}}

\newcommand{\Gapp}{G^\mathrm{c}}

\newcommand{\Gappc}{\Gapp_\fraction}

\newcommand{\Gfail}{\Gapp(\caf)}
\newcommand{\Gfailc}{\Gappc(\caf)}

\newcommand{\Gini}{\Gapp(0)}

\newcommand{\Gloc}{\Gamma^\mathrm{loc}}
\newcommand{\Glocc}{\Gamma^\mathrm{loc}_\fraction}

\newcommand{\Gfull}{\Gamma^\mathrm{perfect}}

\newcommand{\lx}{\ell_\mathrm{x}}

\newcommand{\Lx}{L_\mathrm{x}}
\newcommand{\Ly}{L_\mathrm{y}}
\newcommand{\Lfail}{\caf}

\newcommand{\Lc}{L_\mathrm{c}}
\newcommand{\Lnc}{L_\mathrm{nc}}

\newcommand{\epsmax}{\varepsilon_\mathrm{max}}

\newcommand{\ca}{a}
\newcommand{\dca}{\Delta \ca}

\newcommand{\xt}{\ca}
\newcommand{\caf}{\ca_\mathrm{f}}

\newcommand{\dGini}{\dgap(0)}
\newcommand{\dGfail}{\dgap(\caf)}
\newcommand{\hatdGloc}{\hat\gamma^\mathrm{loc}}
\newcommand{\dGloc}{\gamma^\mathrm{loc}}
\newcommand{\dGinic}{\dgapc(0)}
\newcommand{\dGfailc}{\dgapc(\caf)}
\newcommand{\dGlocc}{\gamma^\mathrm{loc}_\fraction}
\newcommand{\dG}{\dgap}
\newcommand{\dgap}{g^\mathrm{c}}
\newcommand{\dgapc}{g^\mathrm{c}_\fraction}

\newcommand{\Pcum}{\mathcal{P}}

\newcommand{\pdf}{p}

\newcommand{\gs}{g^\star} %

\newcommand{\sone}{s_1}
\newcommand{\Nw}{N_\mathrm{w}}

\newcommand{\Pcummaster}{\mathscr{P}}
\newcommand{\Pcummasterloc}{\Pcummaster_\mathrm{loc}}

\newcommand{\CDF}[1]{\Pcum[{#1}]}
\newcommand{\PDF}[1]{\pdf[{#1}]}

\newcommand{\GumbelCDF}{\mathscr{G}}

\begin{document}

\title{\papertitle}
\author{Antoine Sanner}
\email{asanner@ethz.ch}
\affiliation{Institute for Building Materials, ETH Zurich, Switzerland}

\author{Luca Michel}
\affiliation{Institute for Building Materials, ETH Zurich, Switzerland}

\author{David S. Kammer}
\email{dkammer@ethz.ch}
\affiliation{Institute for Building Materials, ETH Zurich, Switzerland}

 \date{\today}

\myabstract{
Disordered spring networks are a well-established model system to study fracture in a wide range of materials, from ceramics to polymer networks and mechanical metamaterials, across length scales from the atomistic to the macroscopic.
A central quantity characterizing fracture is the apparent fracture energy $\Gapp$, which measures the resistance to the propagation of a preexisting dominant crack.
While it is well established that disorder can increase $\Gapp$ through crack arrest by local inhomogeneities, its dependence on the degree of disorder remains poorly understood.
Here, we study the effect of varying concentrations of missing bonds on crack propagation of an otherwise perfect two-dimensional triangular network of springs. 
For a given network with a fixed concentration of missing bonds, the apparent fracture energy $\Gapp(\ca)$ increases with crack advance $\ca$. 
This behavior can be explained by mapping the effect of the missing bonds onto an equivalent local fracture energy landscape $\Gloc(\ca)$ and applying established theories linking planar crack arrest with fluctuations in $\Gloc(\ca)$.
For increasing fraction of missing bonds $\nu$, the standard deviation of the fluctuations of $\Gloc$ increases with $\sqrt{\fraction}$, which we explain by considering a random-walk-like superposition of perturbations caused by individual missing bonds. We demonstrate that as a consequence of crack arrest by fluctuations in $\Gloc$, the average $\Gapp(\ca)$ follows the same $\sqrt{\fraction}$ scaling.
Furthermore, we observe that the probability density of $\Gloc$ has an exponential tail leading to a logarithmic increase of $\Gapp(\xt)$ with crack advance $\xt$. 
Our results quantitatively link microstructural disorder to macroscopic fracture energy and paves the way for quantitative predictions of the fracture energy in a wide variety of materials.
}

\mykeywords{crack-pinning, fracture toughness, disorder, spring network}

\maketitleandabstract

\section{Introduction}

Defects and disorder are ubiquitous in materials and profoundly influence their resistance to fracture. The nature of the dominant defects, however, varies widely across material classes. Microscopic voids govern the fracture properties in brittle materials, such as glasses and ceramics ~\cite{griffith_vi_1920,lawn_fracture_1993,curtin_microcrack_1990}; manufacturing imperfections often dominate in the case of mechanical metamaterials ~\cite{bhuwal_discovery_2023,zaiser_disordered_2023,romijn_fracture_2007,karapiperis_prediction_2023,fulco_disorder_2025}; while the heterogeneity of polymer networks typically controls fracture in the case of rubber-like materials~\cite{tauber_stretchy_2022,arora_fracture_2020,sakai_highly_2010,yang_polyacrylamide_2019,barney_fracture_2022}. Understanding how disorder influences fracture resistance is not only of fundamental interest but also has practical implications for the design of fracture-resistant materials. To study these effects in a controlled setting and gain fundamental insights relevant across a range of materials and scales, spring network models can be used. In such models, defects and disorder can be introduced via missing or weakened bonds in lattices~\cite{curtin_microcrack_1990,urabe_fracture_2010,hartquist_fracture_2024,broedersz_criticality_2011}, random bond strengths~\cite{kahng_electrical_1988,shekhawat_damage_2013}, or amorphous network structures~\cite{arora_coarsegrained_2022}. 

A central question in this context is how the presence and density of defects govern the competition between crack nucleation and crack propagation, and whether disorder ultimately toughens or weakens a material. Using spring network-like models, it has been established that in the absence of a preexisting crack, defects promote crack nucleation, and thus lower fracture resistance ~\cite{alava_statistical_2006,shekhawat_damage_2013}. When a dominant crack is already present, failure depends on how the defects affect the propagation of this dominating crack, which leads to a distinct and more complex dependence of failure on the presence of defects. When the defect density is large, the material is weakened by further increasing defect densities~\cite{hartquist_fracture_2024,fulco_disorder_2025,chouzouris_how_2026}, following the same trend as in the absence of the dominating crack. At low defect densities in precracked systems, however, increasing disorder may enhance resistance to crack propagation and thereby increase the ultimate stress of the material~\cite{curtin_microcrack_1990,urabe_fracture_2010,chouzouris_how_2026}.  

This disorder-induced resistance to the propagation of a dominant crack can be quantified by the apparent fracture energy  
$G^c$,  where we adopt the term \enquote{apparent} to emphasize that  
$G^c$ differs from the surface energy in the ideal equilibrium case considered by Griffith~\cite{griffith_vi_1920}. In heterogeneous brittle materials,  
$G^c$ typically increases with crack advance $a$ because the crack is pinned by microstructural heterogeneity~\cite{kendall_control_1975,charles_crack_2002,hossain_effective_2014,lebihain_effective_2021,sanner_why_2024,sanner_less_2025,curtin_microcrack_1990},  
and it therefore ultimately exceeds the ideal value. Quantitative predictions of the toughening caused by crack pinning have been obtained in the case of a heterogeneous landscape of \emph{local} fracture energy, $\Gloc$~\cite{curtin_microcrack_1990,charles_crack_2002,roux_effective_2003,roux_selfconsistent_2008,patinet_quantitative_2013,demery_microstructural_2014}. However, a similar quantitative framework for toughness arising from elastic heterogeneity, such as missing bonds, remains incomplete. A promising approach is to map the elastic heterogeneity onto an equivalent $\Gloc$ landscape and apply planar crack pinning theories~\cite{joanny_model_1984,sanner_why_2024}, which have been partially explored in triangular spring networks~\cite{curtin_microcrack_1990}. However, a theoretical understanding of how fluctuations in $\Gloc$ depend on defect density is still lacking.

Here, we use a spring-network model to investigate how multiple randomly placed missing bonds affect the propagation of a preexisting crack in a triangular spring network. We show that for a given elastic heterogeneity induced by a fraction of missing bonds $\fraction$, the standard deviation of the equivalent $\Gloc$ grows as $\sqrt{\fraction}$. We explain this scaling based on the random-walk-like superposition of the perturbations induced by individual missing bonds. These enhanced local fluctuations lead to a \emph{systematic} increase in the apparent fracture energy $\Gapp(a)$ for large crack advances $a$, because crack arrest is governed by the largest values of $\Gloc$ encountered along the crack path. 
Our results establish the local mechanisms through which disorder enhances fracture resistance and provide a quantitative framework for connecting microscopic defect statistics to macroscopic toughness.

\section{Problem statement and methods} \label{sec:methods}

\begin{figure}[htb]
    \centering
    \includegraphics{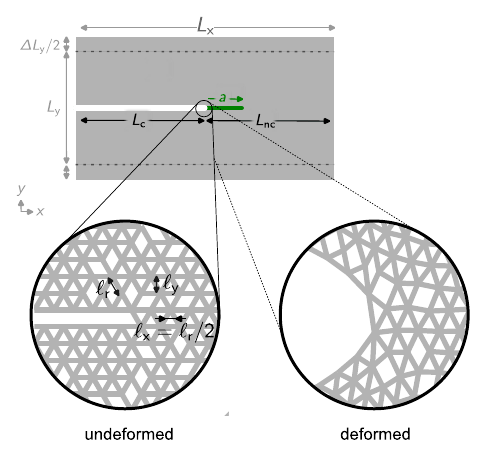}
    \caption{\textbf{Pre-cracked spring network with randomly placed missing bonds.} 
    All dimensions are given in the undeformed state, where all springs are at their rest length $\ell_r$. 
    The deformed configuration is shown at the onset of crack propagation, when the spring at the crack tip reaches its maximum length $(1 + \varepsilon_{\max})\ell_r$ with $\varepsilon_{\max} = 1$. 
    The green line symbolizes crack growth by a length $a$.}
    \label{fig:setup}
\end{figure}

In this work, we study fracture in a triangular network of linear springs under quasi-static loading (Fig.~\ref{fig:setup}). 
In the undeformed state, all springs have rest length $\ell_r$ and carry no initial force. 
When deformed, the force in a spring is
\begin{equation}
f = k(\ell - \ell_r),    
\end{equation}
where $\ell$ is the deformed length and $k$ the spring stiffness. 
A spring breaks abruptly when its strain $\varepsilon = \ell/\ell_r - 1$ exceeds the threshold $\varepsilon_{\max} = 1$. 
This choice promotes straight crack propagation while keeping the simulation domain moderate in size, in line with our prior work~\cite{sanner_less_2025}.

For simulations dedicated exclusively to crack initiation, we consider a domain of width $L_x = 200\,\ell_r$ and height $L_y = 100\,\sqrt{3}/2\,\ell_r$ (\ie{}, 200 $\times$ 100 unit cells), containing a pre-crack of length $L_c = 100\,\ell_r$. 
To analyze crack propagation, we consider a system with dimensions $L_x = 800\,\ell_r$, $L_y = 200\,\sqrt{3}/2\,\ell_r$, and $L_c = 160\,\ell_r$. 
These dimensions are chosen to minimize boundary effects~\cite{deng_nonlocal_2023,rivlin_rupture_1953,kermode_lowspeed_2008,long_fracture_2016}, and we demonstrate that our result are not significantly affected by the system size in appendices~\ref{app:convergence}~and~\ref{app:convergence-Gini}. 
We impose uniform vertical displacements on the top and bottom boundaries while constraining all lateral displacements of the boundaries to zero.

Loading proceeds in increments of vertical displacement $\Delta L_y$. 
After each increment, we relax the network to static equilibrium using the FIRE minimization algorithm implemented in LAMMPS~\cite{bitzek_structural_2006,guenole_assessment_2020,thompson_lammps_2022}. 
Springs that exceed $\varepsilon_{\max}$ are removed one at a time, starting from the most stretched, following~\citet{dussi_athermal_2020}. 
After each removal, equilibrium is recomputed, and the process is repeated until no further springs break.
The next displacement increment is then applied.

We define the apparent fracture energy $\Gapp(a)$ as the critical applied elastic energy release rate, $G$, required to propagate the crack from its initial length $L_c$ to $L_c + a$. 
In practice, we compute $G$ as the elastic energy density ahead of the crack tip in the region $x \in [3L_x/4, L_x]$, multiplied by the strip height $L_y$. 
This measure is independent of crack length and becomes exact in the limit $L_c / L_y \to \infty$ and $L_x / L_y \to \infty$ due to translational symmetry. 
Because of the independence on the crack length, prescribing vertical displacement is equivalent to imposing an energy release rate.

\section{Results: Randomly removing bonds increases the fracture energy for failure} \label{sec:random}

\begin{figure}
    \centering
    \includegraphics{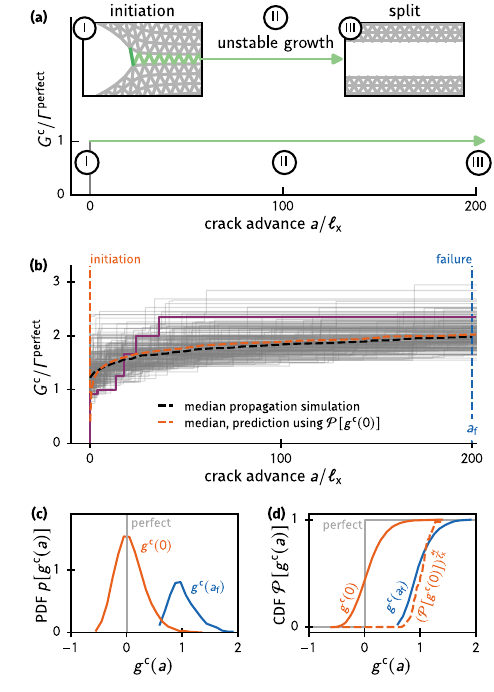}
    \caption{\textbf{Removing bonds from a perfect network increases the apparent fracture energy.}
Apparent fracture energy, $\Gapp$, as a function of crack advance $\xt$ for (a) a pre-cracked but otherwise perfect network,
        and (b) imperfect pre-cracked networks, where a fraction $\fraction=0.2$ of bonds are missing. We normalize $\Gapp$ using the fracture energy of the perfect network $\Gfull$.
        The gray lines in panel (b) show the response of 200 different realizations of the random network, and the purple line highlights one particular realization. 
        Note that these curves do not include the complete range from initiation to complete splitting of the two networks and focus on the range $\ca<\caf=200\,\lx$. 
        Panel (c) shows the corresponding probability density functions (PDFs) of the normalized deviation in fracture energy $\dG = (\Gapp -\Gfull) / \Gfull$ required for the initiation of crack growth, $\dGini$, and for failure, $\dGfail$.        
        The corresponding cumulative distribution functions (CDFs) $\CDF{\dG}$ are shown in panel (d), 
        and the dashed line shows the prediction of the failure distribution based on the initiation distribution using Eq.~\ref{eq:PG}, see Sec.~\ref{sec:statistical-distributions}. 
        In panel (b), we also compare  the evolution of the median of $\Gapp(\xt)$ predicted by the same equation (orange dashed line) and that extracted from the ensemble of crack propagation simulations (black dashed line).
        The crack initiation distributions are computed based on an ensemble of 10000 simulations of the onset of crack propagation, 
        and the failure distributions are based on an ensemble of $\simeq 250$ simulations of crack propagation. 
    }
    \label{fig:rcurve-example}
\end{figure}

We begin by considering a reference case: a pre-cracked but otherwise perfect triangular spring network. 
We gradually increase the driving force for crack propagation, \ie{}, the elastic energy release rate $G$, by increasing the prescribed vertical displacement at the boundaries.
The first bond to break is at the crack tip (Fig.~\ref{fig:rcurve-example}a-I). In this perfect system, this single event triggers the instantaneous failure of the entire network along a perfectly straight fracture path (Fig.~\ref{fig:rcurve-example}a-III). This sudden failure occurs at a constant energy release rate (Fig.~\ref{fig:rcurve-example}a), meaning that the fracture resistance is characterized by a single value: the critical energy release rate at initiation of propagation, which we denote as the fracture energy $\Gfull$, where the superscript refers to the $\emph{perfect}$ network. This value serves as a reference for the imperfect networks studied below.

Next, we consider networks with a fraction $\fraction=0.2$ of randomly removed bonds. 
Fracture response varies significantly between realizations (Fig.~\ref{fig:rcurve-example}b). 
In contrast to the perfect network, rupture of the first bond does not trigger complete failure. Instead, the crack advances intermittently, alternating between unstable growth and arrest, 
and the apparent fracture energy $\Gapp(a)$ (the critical $G$ required to sustain propagation) increases with crack advance $a$. 

Although we analyze the full function $\Gapp(a)$, in this work we primarily refer to two specific values: (i) the initiation fracture energy $\Gapp(0)$, required to break the first bond, and (ii) the failure fracture energy $\Gfail$, where the crack has advanced by the critical length $\caf=200\,\lx$. We choose $\caf < \Lnc$, the length at which the entire strip would be broken, in order to avoid boundary effects, as discussed in Appendix~\ref{app:convergence}.
To highlight deviations from the perfect network, we introduce the normalized quantities 
\begin{equation}
\label{eq:g_norm}
\dgap(\ca) = (\Gapp(\ca)-\Gfull)/\Gfull
\end{equation}
which measure the fracture energy deviation of a given system relative to the perfect network.

The initiation and failure fracture energies are affected in two contrasting ways by the presence of the missing bonds. 
While the initiation fracture energy $\Gini$ remains, on average, close to $\Gfull$, \ie, $\left\langle\dGini\right\rangle\simeq0$, the failure fracture energy is consistently increased, $\Gfail > \Gfull$, \ie{} $\dGfail > 0$, as shown by the normalized probability density functions $\pdf[\dG(\xt)]$ (PDF) and cumulative distributions $\Pcum[\dG(\xt)]$ (CDFs) in Fig.~\ref{fig:rcurve-example}c,d. Although this figure shows only results for $\fraction=0.2$, we obtained qualitatively similar results for missing bond fractions $\nu=0.01, 0.05$ and $0.1$ (results not shown).  
Thus, we see that in the regime of $\nu\leq0.2$ considered in this work, random bond removal systematically toughens the network at failure.
As discussed in the next section, the resulting intermittent crack propagation is a direct signature of crack arrest caused by local heterogeneity.

\section{Local fracture energy fluctuations arrest the crack} \label{sec:missing-bond-fraction}

We now show that the increase in the apparent fracture energy, $\Gapp(\ca)$, with crack advance, $\ca$, originates from crack arrest induced by spatial fluctuations of the local fracture energy, and we derive a theoretical prediction for this increase following Refs.~\cite{curtin_microcrack_1990,charles_crack_2002}.
For simplicity, and only in this section, we restrict the analysis to a straight crack path by allowing bond breaking only along the initial crack plane ($y=0$).
This protocol differs from all other simulations presented in this paper, where the crack path is not prescribed and typically deviates slightly from a straight line.
As shown at the end of this section, predictions assuming a straight path remain accurate even in those cases, indicating that crack arrest is the dominant toughening mechanism in the regime we study here, while crack path roughness plays only a minor role.

\subsection{Local fracture energy $\Gloc$}

We first quantify the landscape of local fracture energy $\Gloc(\xt)$ that governs the arrest of the crack.
To obtain $\Gloc$, we simulate forced crack growth by artificially extending the crack length, rather than letting it propagate spontaneously. For each crack position, the applied displacement is increased gradually until the bond at the crack tip reaches the failure threshold $\epsmax = 1$, and the corresponding elastic energy release rate is taken to define $\Gloc$ at that point. The resulting values of $\Gloc$ fluctuate due to the varying constellations of missing bonds encountered as the crack advances, as illustrated by the light-blue bars in Fig.~\ref{fig:Gloc}.

\subsection{Crack arrest}

\begin{figure}
    \centering
    \includegraphics{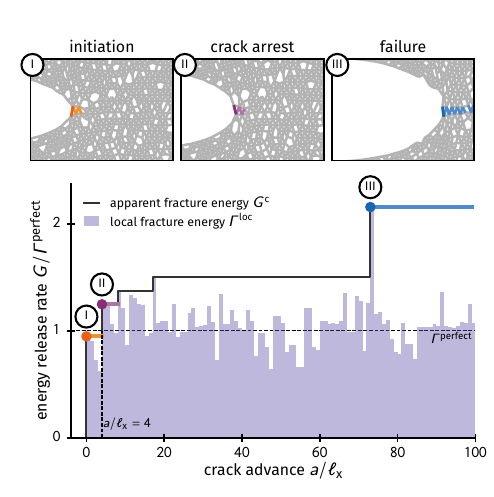}
    \caption{\textbf{The apparent fracture energy $\Gapp(\ca)$ is equal to the strongest local fracture energy encountered along the crack path up to the point $\ca$.}
    Evolution of the local fracture energy $\Gloc$ (light blue bars) and of the apparent fracture energy $\Gapp$ (black line) as a function of crack advance in an imperfect network with a fraction $\fraction=0.1$ of the bonds missing. The apparent fracture energy, $\Gapp(a)$, is defined as the critical applied elastic energy release rate $G$ required to sustain crack propagation at crack advance $a$. Colored segments of the curves indicate unstable crack growth events. (insets I to III) Deformed configurations at the onset of the unstable event corresponding to each label. Bonds broken during the unstable event are highlighted in color, with darker colors indicating those that break first.
    }
    \label{fig:Gloc}
\end{figure}

Next, we return to the case of spontaneous crack propagation and explain how local fluctuations in $\Gloc(\xt)$ lead to $\Gfail \geq \Gini$, and more generally how $\Gapp(\xt)$ is related to $\Gloc(\xt)$. As before, we restrict bond breaking to the initial crack plane to ensure a straight crack path, consistent with the definition of $\Gloc$. We analyze the spontaneous crack propagation that occurs upon the bonds reaching their maximum strain $\epsmax$ as the applied $G$ is gradually increased.

In general, the crack advances only if the applied $G$ is greater than or equal to the local fracture energy $\Gloc$. Thus, from the initial crack position at $\xt = 0$, propagation will occur when $G$ reaches the local fracture energy $\Gloc$ at that point, \ie, $\Gini = \Gloc(0)$ (see Fig.~\ref{fig:Gloc}-I). Considering the realization shown in Fig.~\ref{fig:Gloc}, the local fracture energy at $\xt = \lx$ is slightly lower than that of $\xt = 0$, so the crack continues advancing. The crack keeps propagating to successive positions $\xt = n\lx$ as long as $\Gloc(n\lx) < G$. However, when it encounters a location where $\Gloc$ exceeds $G$\textemdash in the example shown, this occurs at $n=\xt/\lx=4$\textemdash the crack arrests (see Fig.~\ref{fig:Gloc}-II). Further propagation is then only possible if $G$ increases enough to overcome this local peak. This sequence of unstable crack growth and arrest continues as $G$ is gradually increased, with each new arrest requiring the crack to overcome a higher local maximum of $\Gloc$.

Consequently, the largest peak in local fracture energy encountered up to position $m\lx$ determines the applied energy release rate required for propagation, which defines the apparent fracture energy
\begin{equation} \label{eq:GfailfromGini}
\Gapp(m \lx) = \underset{1 \leq n \leq m}{\max}\,\Gloc(n\lx)~.
\end{equation}
In particular, this means that the failure fracture energy $\Gfail$ is always greater than or equal to the initiation fracture energy.
Moreover, for $m = \Lfail / \lx = 200$ it is extremely unlikely that all 200 values of $\Gloc(n\lx)$ lie below the mean $\langle \Gloc \rangle = \Gfull$.
This explains why in our simulations we never observe $\Gfail < \Gfull$ (Fig.~\ref{fig:rcurve-example}c,d).

\subsection{Statistical distribution}
\label{sec:statistical-distributions}

Having established the deterministic relationship between $\Gapp$ and  $\Gloc$ on the example of a specific realization of the network, we use Eq.~\ref{eq:GfailfromGini} to predict the statistical distribution of the normalized apparent fracture energy, $\dgap(\xt)$, for a given value of $\fraction$.
This derivation is particularly simple because the applied elastic energy release rate, $G$, is independent of crack length, and it corresponds to a special case of the more general formulations presented in Refs.~\cite{curtin_microcrack_1990,charles_crack_2002}. 

To predict the statistics of $\dgap(\ca)$, we determine the statistical properties of $\dGloc(\ca) = (\Gloc(\ca)-\Gfull)/\Gfull$ under two simplifying assumptions.
First, we assume statistical translational invariance, meaning that the (cumulative) probability distribution is independent of the crack advance $\xt$, which holds when the crack tip is far from the boundaries. Due to the arbitrary initial crack length considered in this work, this assumption results in the observation that $\Pcum[\dGloc]$ must be equivalent to $\Pcum[\dGini]$, and we note that the latter distribution can be easily obtained from simulations on crack initiation (Fig.~\ref{fig:rcurve-example}d).. Second, we assume that $\dGloc(\xt)$ is spatially uncorrelated, \ie,\ that $\dGloc(n\lx)$ and $\dGloc(m\lx)$ are independent random variables for $n \neq m$. Appendix~\ref{app:gloc-correlation} confirms that correlations between $\dGloc(n\lx)$ and $\dGloc((n+d)\lx)$ are indeed negligible even for $d=1$. 

We now derive the cumulative probability distribution of the energy release rate, $\Pcum[\dgap(m \lx)](\gs) = P(\dgap(m \lx) < \gs)$, \ie, the probability that $\dG(m \lx)$ is smaller than a given value $\gs$.
From Eq.~\ref{eq:GfailfromGini}, the condition $\dG(m\lx)<\gs$ is met only when all $\dGloc(n)$ encountered along the path $1 \leq n \leq m$ are smaller than $\gs$. Since the $\dGloc(n\lx)$ are independent and identically distributed, this occurs with probability
\begin{equation}\label{eq:PG}
    \Pcum[\dG(m \lx)](\gs) = \left( \Pcum[\dGloc](\gs) \right) ^{ m} . 
\end{equation}
Using Eq.~\ref{eq:PG}, and taking the distribution of $\Pcum[\dGloc]$ to be equal to the $\Pcum[\dGini]$ extracted from our simulations, we can make predictions for the median of $\dgap(\xt)$ and the cumulative distribution $\Pcum[\dGfail]$; these predictions are shown as dashed line in Fig.~\ref{fig:rcurve-example}b,d) for missing bond fraction $\fraction=0.2$ and agree closely with the results of the $\simeq 250$ crack propagation simulations.
This agreement is especially remarkable given that the theory assumes a straight crack path, whereas the simulated crack path deviates slightly from straightness (Appendix~\ref{app:crack-path}).
Thus, in the regime considered here, crack arrest by local fluctuations in $\dGloc$ is the dominant toughening mechanism, while crack path roughness plays only a minor role.

However, the predictions shown in Fig.~\ref{fig:rcurve-example}b,d relies on the distribution $\CDF{\dGloc}$, which in the above analysis was taken directly from simulations at a fixed bond fraction $\fraction$.
In the next section, we present a model that predicts the CDF $\CDF{\dGlocc}$ for a given proportion of missing bonds, $\fraction$ (note that from now on we add the subscript $\fraction$ to emphasize this dependence). The dependence of $\CDF{\dGlocc}$ on $\fraction$ is of practical importance since this relationship governs how $\dgapc(a)$ scales with the fraction of removed bonds.

\section{Fluctuations increase with missing bond fraction} \label{sec:c-dependence}

\begin{figure*}
    \centering
    \includegraphics{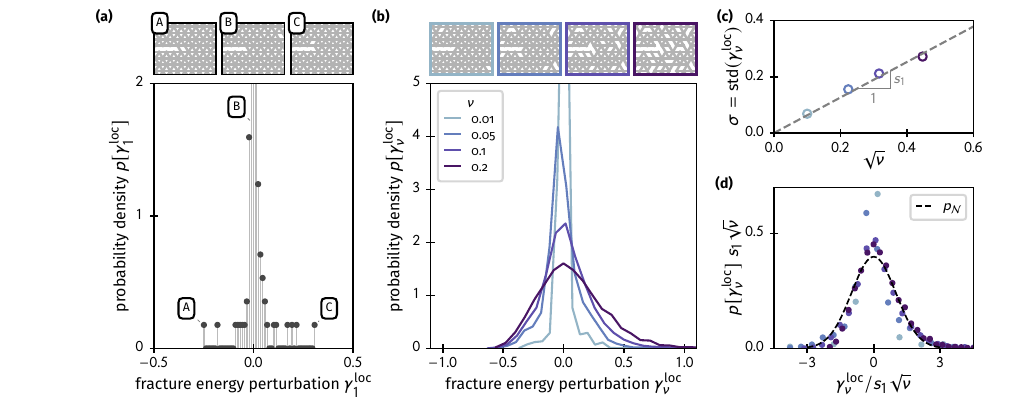}
    \caption{\textbf{The fluctuations of the local toughness increase with the fraction of removed bonds because the perturbations of each removed bond add up.}
    (a) Probability density (PDF) of the local fracture energy for a network with one randomly placed missing bond. 
     We show the normalized deviation from the perfect network value $\dGloc_{1} = (\Gloc_{1} - \Gfull) / \Gfull$.
    The insets show examples of missing bond placements leading to (A) a strong decrease in $\dGloc_{1}$, (B) almost no change in $\dGloc_{1}$, and (C), a strong increase in $\dGloc_{1}$.
        (b) Probability density of $\dGlocc$ for randomly removed bonds for various values of $\fraction$. For each $\fraction$, the distribution is based on an ensemble of 10000 realisations.
        (c) Standard deviation (STD) of $\dGlocc$ as a function of the fraction of removed bonds. The circles are numerical results, and the dashed line is the prediction by Eq.~\ref{eq:std-prediction}. This prediction uses the standard deviation of the distribution for one missing bond $\pdf[\dGloc_1]$.
        (d) Same data as panel b, but normalized by the predicted standard deviation. The dashed line shows a standard Gaussian probability density $p_\mathcal{N}$.
    }
    \label{fig:Gini-void-fraction-dependence}
\end{figure*}

In the previous section, we showed that the failure fracture energy $\dGfailc$ is determined by the largest peak in the local fracture energy $\dGlocc$ encountered along the crack path, and, for a given fraction of missing bonds, we determined the link between the distributions of local fracture energies and apparent fracture energies that induce crack propagation.
Here, we show that the fluctuations in $\dGlocc$, and hence the expected maximum $\dGlocc$ and $\dgapc(\caf)$, increase as $\sqrt{\fraction}$ with the fraction $\fraction$ of missing bonds. 
We rationalize this scaling by first analyzing the perturbation caused by a single missing bond, which we denote $\dGloc_1$, and then showing that the contributions of multiple missing bonds add up stochastically in a random-walk-like manner.

Figure~\ref{fig:Gini-void-fraction-dependence}a shows the probability density function of the change in local fracture energy caused by removing one bond, $\pdf[\dGloc_1]$ (details in Appendix \ref{app:std-independence}).
The probability density $\pdf[\dGloc_1]$ has a sharp peak around $0$ because the influence of a missing bond decays rapidly with distance~\cite{sanner_less_2025}, so that many possible locations lead to a value of $\dGloc_1$ close to $0$.
More importantly, $\pdf[\dGloc_1]$ is nearly symmetric about $0$, indicating that the probability of increasing or decreasing the fracture energy of the system by removing a single bond is comparable.

For a given realization of the system with $M>1$ missing bonds, corresponding to removing a fraction $\fraction=M/N$ of the $N$ bonds initially present in the network, we can approximate the total fracture energy perturbation $\dGloc_\fraction$ by summing the individual contribution of each missing bond in that realization, \ie,  
\begin{equation} 
\dGloc_\fraction = \sum_{i=1}^{M = \fraction N} \dGloc_{1,i}, 
\end{equation} 
where each random variable $\dGloc_{1,i}$ is independent and independently distributed according to the single-bond distribution $\pdf[\dGloc_1]$.

The perturbations induced by each removed bond are stochastic and accumulate as the number of removed bonds increases, such that $\dGloc_\fraction$ is a random variable with $\pdf[\dGloc_\fraction]$ that broadens in the manner of a random walk, which is indeed what we observe in our simulations, see Fig.~\ref{fig:Gini-void-fraction-dependence}b.
More quantitatively, the variance of $\pdf[\dGloc_\fraction]$ is expected to grow linearly with $M$, because $\dGloc_\fraction$ is the sum of $M$ independent random variables.
Consequently, the standard deviation of $\dGloc_\fraction$ will be given by
\begin{equation}
\sigma(M) = \sqrt{M}\,\sigma_1,
\end{equation}
where $\sigma_1$ is the standard deviation of the single-bond distribution $\pdf[\dGloc_1]$. 
While it seems that $\sigma$ depends on the system size, $N$, because $M = \fraction N$, this effect is canceled out by the fact that $\sigma_1 \propto 1 / \sqrt{N}$, 
see Appendix \ref{app:std-independence}. Hence the corresponding system-size independent expression is 
\begin{equation}
\label{eq:std-prediction}
\sigma(\fraction) = \sone \sqrt{\fraction}.
\end{equation}
The constant $\sone = \sigma_1 \sqrt{N}$ captures the fluctuations induced by a single bond removal and, in contrast to $\sigma_1$,  is independent of the system size.
This prediction is in excellent agreement with the result from brute force simulations of $\dGlocc$, see Fig.~\ref{fig:Gini-void-fraction-dependence}c, 
indicating that treating the perturbations from each removed bond as independent is a good approximation of the actual behavior of the system.
Further support for this assumption is presented in Appendix~\ref{app:std-independence}, where we show that the entire distribution obtained by superposition of single-bond effects matches closely that obtained via direct simulations of networks with randomly placed missing bonds.

\begin{figure}
    \centering
    \includegraphics{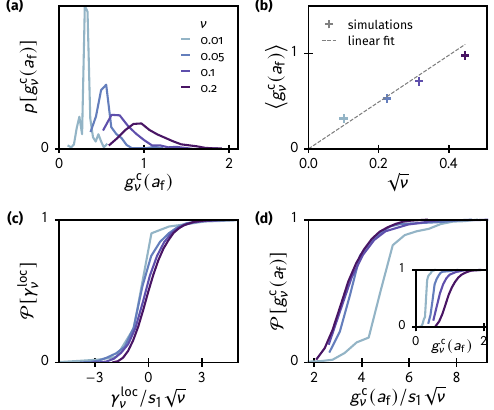}
    \caption{\textbf{The failure fracture energy scales with $\sqrt{\fraction}$.}
    \textbf{(a)} Probability density of the (normalized) failure fracture energy, $\dGfailc$, for different fractions of missing bonds, $\fraction$.
    \textbf{(b)} Average failure fracture energy $\left< \dGfailc \right>$ as a function of the fraction of missing bonds $\fraction$. The dashed line is a linear fit to the simulation results.
    \textbf{(c)} Cumulative distribution of $\dGlocc$, where the $x$-axis is normalized by the predicted standard deviation of $\dGlocc$ $\sigma(\fraction) = \sone \sqrt{\fraction}$.
    \textbf{(d)} Cumulative distribution of $\dGfailc$, where the $x$-axis is normalized by $\sone \sqrt{\fraction}$. The inset shows the same data without normalization.}
    \label{fig:Gfail-void-fraction-dependence}
 \end{figure}

Since $\dGfailc$ is governed by the largest peaks of $\dGlocc$, this broadening of $\pdf[\dGlocc]$ leads to an increase in $\dGfailc$, as confirmed by our simulations (Fig.~\ref{fig:Gfail-void-fraction-dependence}a).
Indeed, given that $\dGfailc$ and $\dGlocc$ are linked through Eq.~\ref{eq:PG}, we would expect that the distribution $\Pcum[\dgapc(\caf)]$ inherits the same $\sqrt{\fraction}$ dependence as $\dGlocc$. This hypothesis is supported by the fact that the (ensemble) average, $\left<\dGfailc\right>$, scales as $\sqrt{\fraction}$, see Fig.~\ref{fig:Gfail-void-fraction-dependence}b.  Note that we use the brackets $\left<\right>$ to denote ensemble averaging.
To state this dependence more formally we use the fact that the rescaled local fracture energy $\phi  =  \dGloc_\fraction  / \sigma(\fraction)$ collapses the cumulative distributions of the local fracture energy $\Pcum[\dGloc_\fraction](\phi \, \sigma(\fraction)) = P(\dGlocc < \phi \, \sigma(\fraction))$ onto a $\fraction$-independent master curve $\Pcummasterloc(\phi)$
\begin{equation} 
\label{eq:glocmaster} 
\Pcum[\dGlocc](\phi \, \sigma(\fraction))  \simeq \Pcummasterloc(\phi), 
\end{equation}
as shown in Fig.~\ref{fig:Gfail-void-fraction-dependence}c and previously shown for the probability density in Fig.~\ref{fig:Gini-void-fraction-dependence}d. Note that we do not show $\Pcummasterloc$ in the figure because we do not make any assumption on its functional form. 
Substituting Eq.~\ref{eq:glocmaster} into Eq.~\ref{eq:PG} indicates that the CDF of the apparent fracture energy should obey 
\begin{equation} 
    \label{eq:gfailmaster} 
\Pcum[\dgapc(\ca)](\phi \, \sigma(\fraction)) \simeq \left(\Pcummasterloc(\phi)\right)^{\ca/\lx}. 
\end{equation} 
which, for a given system size and crack propagation length, depends only on the rescaled fracture energy $\phi$ and is independent of $\fraction$. Thus, Eq.~\ref{eq:gfailmaster} predicts that $\Pcum[\dGfail]$ can be collapsed onto the $\fraction$-independent master curve 
\begin{equation} 
\label{eq:gmaster} 
\Pcum[\dGfailc](\phi \, \sigma(\fraction)) \simeq \Pcummaster_{\dgap}(\phi, \caf).  
\end{equation} 
This prediction is confirmed by our numerical results for $\fraction \geq 0.05$ (Fig.~\ref{fig:Gfail-void-fraction-dependence}d). 
In conclusion, the random-walk–like superposition of single-bond perturbations explains why $\left<\dGfailc\right>$ increases with $\sqrt{\fraction}$, 
showing that toughening arises as a direct consequence of stochastic accumulation of local effects.

\section{Distributions of $\dGlocc$ and $\dGfailc$} \label{sec:exp-tail-log-rise}

\begin{figure*}
    \centering
    \includegraphics{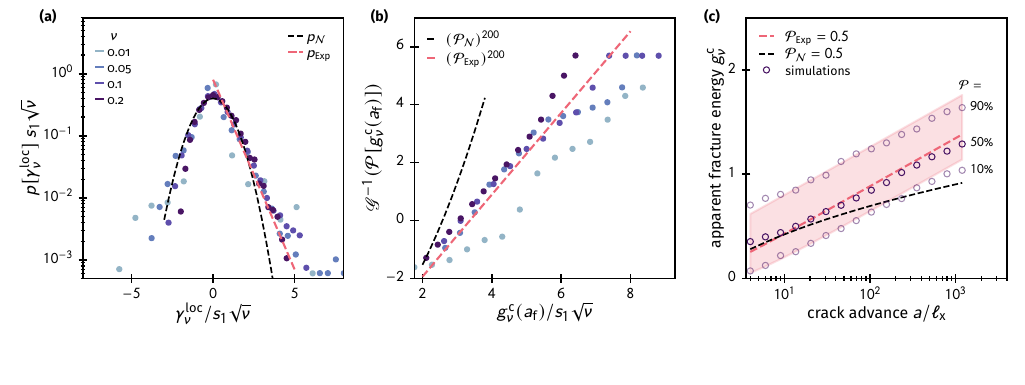}
    \caption{ \textbf{The exponential tail of $\dGlocc$ leads to a logarithmic increase of $\dgapc$ with crack advance.}
        \textbf{(a)} Probability density function of $\dGlocc$ rescaled by, $\sone \sqrt{\fraction}$, the predicted standard deviation (Eq.~\ref{eq:std-prediction}) for different values of $\fraction$. Note that some data points at high $\dGlocc$ are outside of the plotting range. This plot shows the same data as Fig.~\ref{fig:Gini-void-fraction-dependence}d but on a semi-logarithmic scale, so that the standard Gaussian distribution takes the form of a parabola (black dashed) and the exponential distribution a straight line (red dashed line). The exponential distribution is fitted to the numerical data as described in Appendix~\ref{app:exp-tail-fit}. 
        \textbf{(b)} Cumulative distribution of $\dGfailc$ rescaled by the inverse cumulative Gumbel distribution $\GumbelCDF^{-1}(\phi) = - \ln (-\ln \phi)$, where the apparent fracture on the $x$-axis is rescaled by the predicted standard deviation of $\dGlocc$, $\sone \sqrt{\fraction}$. The black and red dashed lines show the result of using the Gaussian and the exponential fits in Eq.~\ref{eq:PG}, respectively.
        \textbf{(c)} Normalized applied energy release rate, $\dG$, as a function of crack advance for $\fraction=0.1$, on a logarithmic scale so that $\ln \xt / \lx$ appears as a straight line. 
        The center dashed line is the median predicted by the exponential fit, Eq.~\ref{eq:masterGumbel}, and the shaded area shows the $10\%$ and $90\%$ quantiles.
        The circles indicate the median and the corresponding quantiles from the crack propagation simulations. 
        The black dashed line is the result of using the Gaussian fit in Eq.~\ref{eq:PG}. 
    }
    \label{fig:mastercurve}
\end{figure*}

We now analyze the functional form of the distributions of $\dGlocc$ and $\dGfailc$ in more detail; we then use these distributions to provide insight into how $\dgapc(a)$ evolves with crack advance $\xt$.
As in the previous section, we approximate $\dGlocc$ as the superposition of many independent perturbations caused by the missing bonds. By the central limit theorem, this suggests that the center of the probability density function $\pdf[\dGlocc]$ should approach a Gaussian form for sufficiently large $\fraction$. We confirm this by rescaling $\dGlocc$ by the prediction for its standard deviation $\sigma(\fraction) = \sone \sqrt{\fraction}$, see Fig.~\ref{fig:mastercurve}a, and using a logarithmic $y$-scale so that a perfect Gaussian appears as an inverse parabola (black dashed line). While the rescaled distributions resembles the Gaussian near the center (for $\fraction \geq 0.05$ at least), the upper tail deviates significantly from the Gaussian and is better described by an exponential form (red dashed line). Such deviations for large values are expected because the underlying single-bond distribution $\pdf[\dGloc_1]$ (Fig.~\ref{fig:Gini-void-fraction-dependence}a) is non-Gaussian. Because the largest peaks in $\Glocc$ arrest the crack, the upper tail of $\PDF{\dGlocc}$ controls the statistics of $\dgapc$ and thus the evolution of $\dgapc(\xt)$ with crack advance.

We now discuss the expected functional form of the cumulative distribution $\Pcum[\dgapc(a)]$ based on the planar crack arrest model and on $\Pcum[\dGlocc]$.
Equation~\ref{eq:GfailfromGini} shows that the value of $\dgapc(\ca)$ is the maximum of the local $\dGlocc$ along the crack path, and thus the  distribution of $\dgapc(\ca)$ should converge to a limiting extreme-value distribution for large crack advances and in particular for $\ca = \caf$. Since $\dGfailc$ is determined by the largest of many values of $\dGlocc$, it is mostly determined by the upper tail of $\PDF{\dGlocc}$, which is described by an exponential. The exponential tail of $\pdf[\dGlocc]$ leads to a Gumbel extreme-value distribution for $\dGfailc$, see Fig.~\ref{fig:mastercurve}b, where we rescale $\Pcum[\dGfailc]$ using the inverse standard Gumbel distribution, $\GumbelCDF^{-1}(\Pcum[\dGfailc](\gs)) = -\ln(-\ln\{\Pcum[\dGfailc](\gs)\})$ so that a perfect Gumbel distribution appears as a straight line. 
To confirm that the exponential behavior of the tail of the distribution determines $\dGfailc$, rather the Gaussian-like center, we also show the predictions of $\CDF{\dGfailc}$ obtained by inserting Gaussian and exponential functional form for $\dGlocc$ into Eq.~\ref{eq:PG} (black and red dashed lines in Fig.~\ref{fig:mastercurve}b). The results obtained considering the exponential tail of $\pdf[\dGlocc]$ match the simulations well, while assuming a Gaussian would lead to underestimate $\dGfailc$ for a given value of cumulative probability. This confirms that $\dGfailc$ is governed by the extreme values of $\dGlocc$, where the Gaussian approximation fails but the exponential distribution accurately captures the observed behavior.

We now show the implications of the exponential tail of $\pdf[\dGlocc]$ for the evolution of $\dgapc(\xt)$ with crack advance, $\xt$, by deriving analytical expressions for $\Pcum[\dgapc(\xt)]$.
We focus on the large $\dGlocc$ limit, where the probability density function $\pdf[\dGlocc]$ is exponential and the cumulative probability $\Pcum[\dGlocc](\phi)$ is close to one. First, we rewrite Eq.~\ref{eq:PG} as
\begin{equation}
    \Pcum[\dgapc(m \lx)](\gs) = e^{m \ln \left( \Pcum[\dGlocc](\gs) \right)}~,
\end{equation} 
where $m =\ca /\lx$ is the number of broken bonds along the straight path, 
and then use the Taylor expansion of the logarithm of $\Pcum[\dGlocc]$ close to one, yielding
\begin{equation}
    \label{eq:PG-taylor}
    \Pcum[\dgapc(m \lx)](\gs) \simeq e^{m \left( 1 - \Pcum[\dGlocc](\gs) \right)}~.
\end{equation}
As demonstrated above, for large values of $\dGlocc$, $\Pcum[\dGlocc]$ can be well approximated by  $\Pcum[\dGlocc](\phi \sigma(\fraction)) =  1 - \exp[-(\phi - \phi_0)/s]$, where $\phi_0$ and $s$ are the parameters fitted in Appendix~\ref{app:exp-tail-fit}; inserting this expression into Eq.~(\ref{eq:PG-taylor}) we obtain
\begin{equation}
\label{eq:masterGumbel}
\Pcum[\dgapc(m \lx)](\phi \, \sigma(\fraction)) = e^{- m e^{-(\phi - \phi_0)/s}} = e^{- e^{-(\phi - \mu(m))/s}},
\end{equation}
a Gumbel distribution with mode (the most likely value of the random variable $\phi$)
\begin{equation}
    \label{eq:master-mode}
    \mu(m) = \phi_0 + s \ln (m)~.     
\end{equation} 
Analogous formulas for the mean and quantiles follow directly from Gumbel statistics.
Hence, the median of $\dgapc(\xt)$ is expected to increase logarithmically with crack advance $\xt = m\lx$, in perfect agreement with our crack propagation simulations, see Fig.~\ref{fig:mastercurve}c.
This logarithmic increase is a consequence of the exponential tail; 
by contrast, a Gaussian tail would yield $\dgapc(a) \propto \sqrt{\ln (\xt/\lx)}$, which significantly underestimates the observed growth (dashed black line).
Another implication of Eq.~\ref{eq:masterGumbel} is that the distributions conserve their shape while they shift, a property not shared by all extreme value distributions, such as the Weibull distribution.
This is visible in Fig.~\ref{fig:mastercurve}c, where the 10\% and 90\% quantiles remain parallel to the median.

In summary, the energy release rate $\dgapc(a)$ increases with $\ln \xt$ and follows a Gumbel distribution. 
We showed that this dependence is a direct consequence of extreme value statistics applied to the exponential tail in the probability density of $\dGlocc$.
An open question for future work is to identify the physical origin of the exponential tail, or, even better, to predict it analytically rather than fitting it empirically. 

We can obtain insight into the probable origin of the fat tail of $\dGlocc$ by noticing 
that the variance of the single bond removal distribution, $\pdf[\dGloc_1]$, is dominated by a few extreme values, see Fig.~\ref{fig:Gini-void-fraction-dependence}a. In this sense, the process of randomly removing bonds is analogous to a random walk with discrete steps $(-1, 1)$. Such a random walk leads to a binomial distribution, which has fat tails similar to our data, although we note that the tails in this case are not a perfect exponential. 
This illustrates that the exponential is not necessarily the functional form expected theoretically, but we chose it for our fit in the interest of simplicity and because yields an excellent prediction of the increase of $\left<\dgapc(a)\right>$ with $a$, see Fig.~\ref{fig:mastercurve}c.

\section{Discussion}

In this work, we have simulated the fracture of triangular networks with random missing bonds, 
showing that the applied energy release rate increases with the fraction of missing bonds as $\sqrt{\fraction}$ and with the crack advance as $\ln \xt$. 
We linked this increasing $\dgapc(\xt)$ to a crack arrest mechanism, where the largest peak in \emph{local} fracture energy, $\dGlocc$, determines the apparent fracture energy, $\dgapc(\xt)$.
The probability distribution of $\dgapc(\xt)$ at a given crack advance $\xt$ is thereby linked to the distribution $\pdf[\dGlocc]$ via extreme value statistics, 
and the increase in $\dgapc(\xt)$ with $\sqrt{\fraction}$ is the direct consequence of the upper tail of $\pdf[\dGlocc]$ scaling with $\sqrt{\fraction}$.
This $\sqrt{\fraction}$ scaling of the fluctuations 
is a result of the stochastic superposition of the perturbation by each removed bond analogous to a random walk. 

The mechanisms of crack arrest and defect-induced perturbations identified here are not specific to the triangular lattice geometry and are therefore expected to extend to other lattice structures and, more generally, to continuum systems, where missing bonds correspond to microvoids or microcracks.
Our results may thus be relevant for a broad class of disordered materials, provided that the defect density remains low and fracture proceeds along a nearly straight crack path.

We discuss below some further implications of our findings and the limitations of our work.

\paragraph{Size-effect}
The logarithmic increase of $\dgapc(\xt)$ with crack advance implies that the fracture energy required to split the sample entirely, $\dgapc(\Lnc)$, has a logarithmic size effect.
That is, for a strip of given height, the critical force to split it grows logarithmically with the uncracked length $\Lnc$. 
This behavior contrasts with the size effect in disordered materials without a pre-existing crack, where the critical stress decreases with system size~\cite{weibull_statistical_1939,urabe_fracture_2010,alava_statistical_2006}.  
The difference arises because, with a pre-existing crack, defects arrest the crack and enhance toughness, whereas failure without a crack is controlled by nucleation at the largest microcracks.

\paragraph{Existence of a maximum fracture energy}

Since $\dGlocc$ is mainly influenced by missing bonds near the crack tip, one might expect that there exist an upper bound for the maximal possible $\dGlocc$, corresponding to the constellation of missing bonds around the crack tip that is optimal for toughening.
If such a bound exists, it is expected to be found after sufficiently long crack advance, \ie{}, $\Gappc(a)$ would saturate for large $a$ instead of increasing logarithmically without bounds. 
We did not observe such a saturation in our simulations (Figs.~\ref{fig:mastercurve}a,c), meaning that this upper bound either does not exist or can only be reached after even longer crack propagation. If a sample was large enough for $\Gappc(a)$ to reach saturation, the size-effect mentioned in the previous paragraph would not exist and the saturated value could be considered a material property. 

\paragraph{Limitations of the theory}

The main simplifying assumptions in our theory are (i) that the crack propagation is planar, (ii) that the values of $\dGlocc$ are uncorrelated along the crack propagation distance, and, (iii) that perturbations of individual missing bonds are independent and superpose linearly into the total local fracture energy, $\dGlocc$ .

First, we discuss the assumption of planar crack propagation. In reality, the crack path is rarely perfectly straight, exhibiting occasional switches to the neighboring plane for small fractions of missing bonds, and a more pronounced roughness for larger fractions of missing bonds, see Fig.~\ref{fig:sup_rough_crack_paths} in Appendix~\ref{app:crack-path}.
One may expect that crack path roughness increases the fracture energy beyond the prediction of our planar crack arrest model due to several mechanisms 
such as: (i) the increase in number of broken bonds due to the increased arclength of the rough crack path~\cite{urabe_fracture_2010,zhang_fiber_2017,persson_effect_2001}, (ii) mode mixity due to misorientation of the crack~\cite{mirkhalaf_overcoming_2014,faber_crack_1983,faber_crack_1983a}, and (iii), unbroken bonds bridging accross the crack faces~\cite{bower_threedimensional_1991,mirkhalaf_overcoming_2014}. However, we argue below that in our case, these mechanisms are negligible compared to the crack arrest by local heterogeneity.

We can completely exclude the increase in number of broken bonds mechanism, 
because the number of broken bonds actually $\emph{decreases}$ as the result of removing bonds from the network, see Fig.~\ref{fig:sup_broken_per_advance} in Appendix~\ref{app:nb-bonds}. 
The typical argument linking fracture energy to the number of broken bonds is based on energy conservation, which does not hold here. The process of crack propagation is unstable, dissipating energy, and is governed by arrest by strong obstacles.

Concerning the bridging chains mechanism, we found in a previous study with a single missing bond that toughening by bridging bonds plays an important role for critical strains of the bonds $\epsmax < 0.3$. For the current study, we deliberately chose $\epsmax = 1$ to avoid these effects and reduce the roughness of the crack path. We can hence expect stronger deviations between the planar crack arrest theory and the simulations for smaller $\epsmax$, which needs to be verified in future work. The regime of small failure strain is for example especially important for brittle architected materials~\cite{quintana-alonso_fracture_2009,lingua_breaking_2026} and for ceramics~\cite{curtin_microcrack_1990}. 

Overall, good match between our theory and our numerical results suggests that in our case the crack arrest by local elastic heterogeneity described by our theory dominates compared to all those mechanisms.

The second important assumption in the crack arrest theory is that $\dGlocc$ is uncorrelated from one crack tip position to the next. This assumption is in fact very good, since the correlation of $\dGlocc$ decays below 0.5 at a distance of one bond and is nearly zero at a distance of two bonds, see Appendix~\ref{app:gloc-correlation}. This small correlation length suggests that the local fracture energy is dominated by a few defects very close to the crack tip.

The third important approximation in our theory concerns the prediction of the standard deviation  of $\dGlocc$, $\sigma(\fraction)$. 
We computed the perturbations caused by removing one bond at different positions from a perfect network  
and added them up (stochastically) to obtain the total perturbation $\dGlocc$. In reality, when one bond has already been removed, the elasticity of the network has changed and hence the response to the second removed bond will be different. While this effect is negligible when the two removed bonds are far apart, it makes a noticeable difference in some cases where the bonds are close to each other and close to the crack tip. This effect becomes increasingly important at high $\fraction$, where it becomes likely that two neighbouring bonds are missing and furthermore the overall elasticity of the network changes significantly. 
Nevertheless, the distributions computed using the assumption of linear independent superposition of single bond perturbations are in excellent agreement with the result of direct simulations, see Appendix~\ref{app:superposition}.

\paragraph{Broader range of fraction of missing bonds} 
In this paper, we focused on the relatively narrow range of missing bond fractions $ \fraction\leq 0.2 $, and we expect both the planar crack arrest model and our prediction of the fluctuations of $\dGinic$ to break down for large $\fraction$. 
A similar study that introduced weak bonds instead of missing bonds~\cite{hartquist_fracture_2024} observed that after an initial increase with the fraction of weak bonds, 
the fracture energy eventually decreases, so that very heterogeneous networks end up weaker than the perfect network. 
We attribute this opposite trend at large disorder to a different failure mode, where bonds break not only at the crack tip, but instead breakage occurs across a larger region around the crack tip. Accordingly, completely different modeling approaches, such as percolation models, were used to predict the toughness in this delocalized failure regime~\cite{arora_fracture_2020,hartquist_fracture_2024}.

While our numerical simulations could in principle probe the $\fraction>0.2$ regime, we did not manage to reach system size convergence at these concentrations, preventing us from determining the \emph{intrinsic} crack resistance curve $\dgapc(\xt)$. 
Missing bond fractions $\fraction>0.2$ require larger system sizes, 
possibly due to approaching rigidity transition at $\fraction=1/3$, 
where the lengthscale of stress heterogeneity diverges ~\cite{maxwell_calculation_1864,hecke_jamming_2009,broedersz_criticality_2011,zhang_fiber_2017,dussi_athermal_2020}.

\paragraph{Generalization to three dimensions} 
The situation in three dimensions is qualitatively different because the crack tip is a line rather than a point. The strong sensitivity to fluctuations in local fracture energy that we observed here are inherent to the fact that the crack tip is a point,
so that at a given instant of crack propagation, crack growth is determined by a single value of the local toughness.
Hence, a single extreme value of local toughness on the crack path is enough to lead to a large failure fracture energy.
The picture is different in three dimensions, where the crack tip is a line, so that the crack growth is determined by the average of the toughness encountered along the crack front. This averaging along the crack front attenuates the fluctuations, and in the limit of a straight crack front interacting with many uncorrelated heterogeneities, $\dgapc$ would be equal to the average $\dGlocc$. However, in reality fluctuations do not average out completely because the crack front can deform, allowing sections of the front to move independently as if they were independent two-dimensional systems. The result is a partial averaging out of the fluctuations that leads to $\dGfailc \sim \sigma(\fraction)^2 \sim \fraction$, see Refs.~\cite{larkin_pinning_1979,imry_randomfield_1975,robbins_contact_1987,roux_effective_2003,demery_microstructural_2014,sanner_why_2024}, which is smaller and is a different scaling to $\dGfailc \sim \sigma(\fraction) \sim \sqrt{\fraction}$, which we observed here. Three-dimensional network simulations need to be carried out to verify this speculation.

\section{Conclusion} \label{sec:conclusion}

We have shown that randomly removing bonds in a triangular spring network increases the failure fracture energy $\Gfailc$, 
with the increase in $\Gfailc$ being proportional to the square root of the fraction of removed bonds $\sqrt{\fraction}$. 
This increase in toughness is the result of the pinning of the crack by the microstructural heterogeneity created by the missing bonds, 
which can be described by a local fracture energy landscape, $\Glocc(\ca)$. 
Our main finding is that the standard deviation of $\Glocc$ increases with $\sqrt{\fraction}$ as a result of the 
the stochastic superposition of the perturbations caused by each removed bond.
Furthermore, we observe that the probability density of $\Glocc$ has an exponential upper tail, 
which leads to a logarithmic increase of the apparent fracture energy with crack advance.  
Our work paves the way to link fracture behaviour to microstrctural features in a variety of disordered materials.

\section{Acknowledgements}
The authors acknowledge Jan van Dokkum, Mohit Pundir, Matthaios Chouzouris, Leo de Waal, and Marcelo Dias for useful discussions. We thank Daniel Rayneau-Kirkhope for writing assistance. The authors acknowledge the Swiss National Science Foundation for financial support under grant numbers 200343 and 10003776.

\section{CRediT authorship contribution statement}
\textbf{Antoine Sanner}: Conceptualization, Methodology, Investigation, Formal Analysis, Data Curation and Visualisation, Writing -- Original Draft.\\
\textbf{Luca Michel}: Conceptualization, Writing -- Review and Editing\\
\textbf{David S. Kammer}: Conceptualization, Supervision, Formal Analysis, Writing -- Review and Editing, Project administration, Funding acquisition.

\section{Declaration of competing interest}
The authors declare that they have no known competing financial interests or personal relationships that could have appeared to influence the work reported in this paper.

\section{Code Availability}
The code used for the numerical simulations is available on ETH GitLab: \url{https://gitlab.ethz.ch/smec/papers-supp-info/2026/fracture-resistance-like-a-random-walk}

\section{Data Availability}
The generated data has been deposited in the ETH Research Collection database. The link to that database can be found in the gilab repository.

\appendix

\section{System size convergence of the apparent fracture energy $\Gapp(a)$}
\label{app:convergence}

\begin{figure}
    \includegraphics{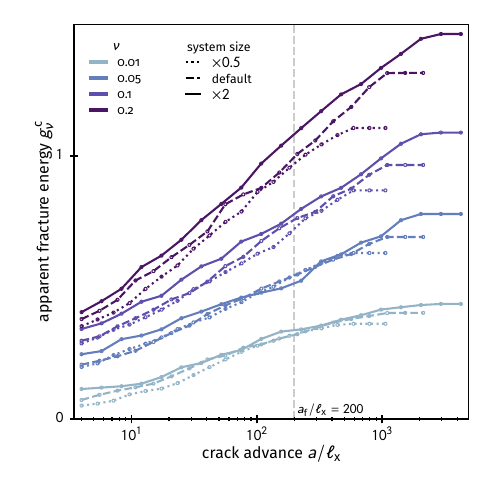}
    \caption{\textbf{System size convergence of the apparent fracture energy $\dgapc(a)$} 
    Ensemble average of the normalized apparent fracture energy as a function of crack advance for different system sizes and void fractions. For each parameter set, we averaged over 100 realizations. The default system size used in the main text is $\Lx= 1600\,\lx$, $\Ly= 400\,\sqrt{3}/2\,\lx$ and $\Lc= 320\,\lx$, and we compare it to systenm sizes twice as small and twice as large.}
    \label{supfig:rcurve_system_size_convergence}
\end{figure}

In this section we show that the apparent fracture energies $\Gapp(\ca)$ reported in the main text Figs.~\ref{fig:rcurve-example}b~and~\ref{fig:mastercurve}c are characteristics of the network that are independent of the system size, \ie{}, independent of the specific choice of $\Lx$, $\Ly$ and $\Lc$. 
Specifically, we show how the ensemble-averaged $\Gapp(\xt)$ changes when halving and doubling all system dimensions compared to the values $\Lx=1600\,\lx$, $\Ly=400\,\sqrt{3}/2\,\lx$, $\Lc=320\,\lx$ used in the main text.
For fractions of missing bonds $\fraction=0.01$ and $0.05$, the apparent fracture energies $\Gapp(a)$ are in excellent agreement across all system sizes, with small deviations attributable to statistical fluctuations, see Fig.~\ref{supfig:rcurve_system_size_convergence}. 
For the large fraction of missing bonds $\fraction=0.1$ and $0.2$, larger and systematic deviations appear, but they remain too small to affect the conclusions of this paper.

Systematic differences also appear for large crack advances $\ca$, where $\Gapp(\ca)$ saturates as $\ca$ approaches $\Lnc$, and this occurs earlier for smaller system sizes. We attribute this saturation to the fact that our expression to compute the elastic energy release rate $G$ assumes an infinite strip, and hence breaks down when the crack is too close to the boundary. The value of $\caf=200\lx$ that we selected is indicated with a vertical dashed line in Fig.~\ref{supfig:rcurve_system_size_convergence} and is small enough to avoid these boundary effects.

\section{System-size convergence of the initiation and local fracture energy distributions}
\label{app:convergence-Gini}

In this section we provide additional details on the simulations used to determine the local fracture energy probability density $\PDF{\dGlocc}$ and verify that the probability density functions shown in the main text (Fig.~\ref{fig:Gini-void-fraction-dependence}bc and Fig.~\ref{eq:gfailmaster}a) are converged with respect to system size.

In the main text, $\PDF{\dGlocc}$ is used to predict the distribution $\PDF{\dgapc(a)}$ through Eq.~\ref{eq:GfailfromGini}. This prediction is extremely sensitive to the upper tail of $\PDF{\dGlocc}$. Although the distribution $\PDF{\dGlocc} = \PDF{\dGinic}$ can already be extracted from the crack propagation simulations, accurately resolving the upper tail requires larger statistics than the $300$ realizations used to characterize $\PDF{\dgapc(\ca)}$, where we were only interested in the overall trend.

For this reason we determined $\PDF{\dGinic}$ based on a dedicated set of simulations containing $10000$ realizations for each $\fraction$. The computational cost of each simulation is reduced because the simulation is stopped after the failure of the first bond instead of simulating the full crack propagation. Furthermore, because the crack does not propagate in these simulations, the system size can be reduced, particularly in the $x$ direction. We therefore use $\Lx=400\,\lx$, $\Ly=200\,\sqrt{3}/2\,\lx$, and $\Lc=200\,\lx$.

\begin{figure}
    \includegraphics{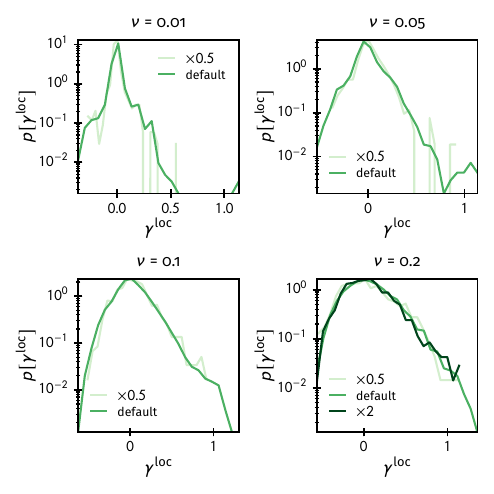}
    \caption{\textbf{System size convergence of the distribution of initiation energy release rate} The base system size used in the main text for the computation of the initiation fracture energy $\dGini$ is $L_x= 400\,\lx$, $L_y= 200\,\lx$ and $L_c= 200\,\lx$. We compare the distribution to systems with different sizes. The number of realizations $n$ is 10000 for the system size used in the main text but we use only 1000 realizations for the other sizes.}
    \label{supfig:system_size_convergence}
\end{figure}

When halving or doubling the system sizes, the probability density functions $\PDF{\dGinic}$ do not change significantly, see Fig.~\ref{supfig:system_size_convergence}, so that the system sizes we chose are appropriate.

\section{$\dGloc$ at different crack tip positions are uncorrelated}
\label{app:gloc-correlation}

\begin{figure}
    \includegraphics{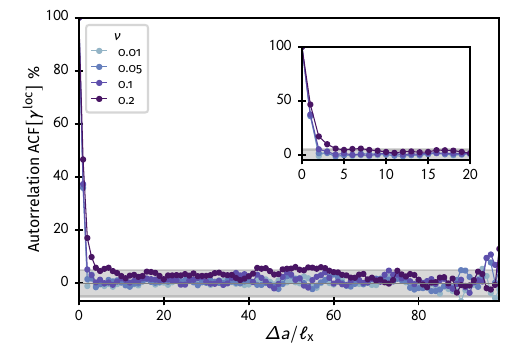}
    \caption{\textbf{$\dGloc(a)$ is uncorrelated with respect to the crack advance} Autocorrelation between the $\dGloc(\ca)$ at different crack tip positions separated by distance $\dca$. The grey shaded area shows correlation between $\pm 5 \%$.}
    \label{supfig:correlation}
\end{figure}

One of the fundamental assumptions underlying the prediction of $\Pcum[\dgap(\ca)]$, Eq.~\ref{eq:PG}, 
is that $\dGloc(\xt)$ is uncorrelated for different crack positions $\xt$. 
Here we test this assumption by computing the autocorrelation of $\dGloc(\xt)$ with $\dGloc(\xt - \dca)$: 
\begin{equation}
    \mathrm{ACF}[\dGloc](\dca) = \frac{\langle (\hatdGloc(\xt)) (\hatdGloc(\xt - \dca)) \rangle}{\langle (\hatdGloc(\xt))^2 \rangle}
\end{equation}
where $\hatdGloc = \dGloc(\xt) - \langle \dGloc \rangle$, and $\langle \cdot \rangle$ denotes the average over all crack positions $\xt$ and over 200 realizations of the disorder. We observe that the strongest correlations occur for the largest fraction of missing bonds $\fraction=0.2$. Even in that case, the ACF is below $50\%$ for a crack advance $\dca$ of only one bond and drops below $5 \%$ as soon as $\dca > 4$, see Fig.~\ref{supfig:correlation}. The drop in correlation is even faster for smaller missing bond fractions. We conclude that the assumption of uncorrelated $\dGloc(\xt)$ is well justified.

\section{Determination of the single bond perturbation probability density distribution $\PDF{\dGloc_{1}}$ and its dependence on the system size}

\label{app:std-independence}

We determine the distribution of the change in local fracture energy $\dGloc_{1}$ caused by removing a single bond from a perfect network.
The system dimensions are the same as used for the simulations of $\dGloc = \dGini$, namely $\Lx=400\,\lx$, $\Ly=200\,\lx$, $\Lc=200\,\lx$, which contains a total number of bonds $N \approx 60000$. Hence, computing $\Gloc(0)$ for each of the $N$ possible choice of missing bond would be extremely costly, and is unnecessary because only missing bonds near the crack tip have a signigicant influence on $\Gloc(0)$. 

We approximate the entire $\pdf[\dGloc_{1}]$ by only computing $\dGloc_{1}$ for the $\Nw \simeq 500$ possible missing bonds within a square window close to the crack tip.  
The histogram of the obtained $\dGloc_{1}$ values yields the probability density of $\dGloc_{1}$ knowing that the missing bond is within the window, which we denote $\pdf[\dGloc_{1 \in \mathrm{w}}]$.

The probability density for the case where the missing bond can be anywhere, $\pdf[\dGloc_{1}]$, is the sum of probability densities for the cases where the missing bond is within the window $\pdf[\dGloc_{1 \in \mathrm{w}}]$ and outside of the window $\pdf[\dGloc_{1 \notin \mathrm{w}}]$, weighted by the respective probabilities $\Nw / N$ and $1 - \Nw / N$ of the missing bond being within or outside of the window:
\begin{equation}
    \label{eq:pdf-dGloc-1-exact}
    \pdf[\dGloc_1](\gs) = \frac{\Nw}{N} \pdf[\dGloc_{1 \in \mathrm{w}}](\gs) + \left(1 - \frac{\Nw}{N}\right) \pdf[\dGloc_{1 \notin \mathrm{w}}](\gs)~.
\end{equation}
If we choose $\Nw$ large enough, the perturbation caused by a missing bond outside of the window is negligible, so that $\pdf[\dGloc_{1 \notin \mathrm{w}}]$, can be approximated by a Dirac distribution, 
\begin{equation}
\label{eq:pdf-dGloc-1-notin-w}
\pdf[\dGloc_{1 \notin \mathrm{w}}](\gs) \simeq \delta(\gs)~.    
\end{equation}
Equations~\ref{eq:pdf-dGloc-1-notin-w}~and~\ref{eq:pdf-dGloc-1-exact} imply that only the removed bonds near the crack tip determine the shape of the tails of $\pdf[\dGloc_{1}](\gs)$.

We now use Eq.~\ref{eq:pdf-dGloc-1-exact} to discuss the implications of the choice of the system size $N$ on $\pdf[\dGloc_1]$ and in particular the standard deviation $\sigma_1$ of $\dGloc_1$. We keep $\Nw$ fixed and furthermore assume that $\pdf[\dGloc_{1 \in \mathrm{w}}]$ is independent of $N$ and hence remains unchanged.
Increasing the system size reduces the probability that the removed bond is within the window, $\Nw / N$, and hence the contribution of $\pdf[\dGloc_{1 \in \mathrm{w}}]$ to $\pdf[\dGloc_1]$. 
For the variance $\sigma_1^2$ of $\pdf[\dGloc_1]$, this implies that 
\begin{equation}
    \label{eq:var-N-dependence}
    \sigma_1^2 = \frac{\Nw}{N} \sigma_{1\in \mathrm{w}}^2~,
\end{equation} 
where we used the additivity of the variance and that the variance of the Dirac distribution is 0.
Hence, 
\begin{equation}
\sigma_1\propto 1 / \sqrt{N}.    
\end{equation}
It follows from Eq.~\ref{eq:var-N-dependence} that $\sone = \sigma_1 \sqrt{N}$ is independent of the system size $N$.

\begin{figure}
    \centering
    \includegraphics[width=0.5\linewidth]{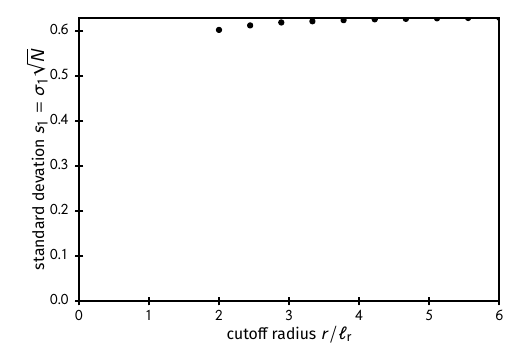}
    \caption{\textbf{The standard deviation of the single bond removal distribution converges with increasing window size}
    Standard deviation of the single bond perturbation $\dGloc_1$ computed using Eq.~\ref{eq:var-N-dependence}
    using a circular window with increasing cutoff radius $r_\mathrm{w}$. Note that $N$ is constant and we show $\sone = \sigma_1 \sqrt{N}$ for a better normalization of the data.
    }
    \label{fig:nw_convergence}
\end{figure}

Finally, we verify that $\sone = \sqrt{\Nw}\,\sigma_{1\in \mathrm{w}}$ converges to a window-size--independent limit as the window used to compute $\dGloc_1$ is enlarged. 
As shown in Fig.~\ref{fig:nw_convergence}, the values of $\sone$ become nearly identical once the window extends beyond approximately $r_\mathrm{w} \simeq 3$. 
This indicates that the perturbations from bonds outside this region are negligible. 
The results presented in the main text are obtained using a square window centered at the crack tip spanning $12 \times 12$ unit cells, corresponding to the largest window shown in Fig.~\ref{fig:nw_convergence}, and are therefore well converged.

\section{Superposition of single-bond perturbations}
\label{app:superposition}

In the main text (Fig.~\ref{fig:Gini-void-fraction-dependence}) we assumed that, in a network with multiple missing bonds, the perturbations of the local fracture energy $\dGlocc$ induced by individual removed bonds superpose independently. Under this assumption, the total perturbation $\dGlocc$ is the sum of the perturbations $\dGloc_{1,i}$ caused by each removed bond $i$. This assumption was used to predict the standard deviation $\sigma$ of $\dGlocc$ as a function of the fraction of missing bonds $\fraction$, see Eq.~\ref{eq:std-prediction}.

Here we further test this superposition assumption by examining whether it also predicts the full probability distribution $\pdf[\dGlocc]$ for networks containing multiple missing bonds. If the perturbations from individual defects are independent, the distribution $\pdf[\dGlocc]$ resulting from the removal of $M=\fraction N$ bonds should be obtained by convolving the single-bond perturbation distribution $\pdf[\dGloc_1]$ (Fig.~\ref{fig:Gini-void-fraction-dependence}a) $M$ times.

\begin{figure*}
    \centering
    \includegraphics{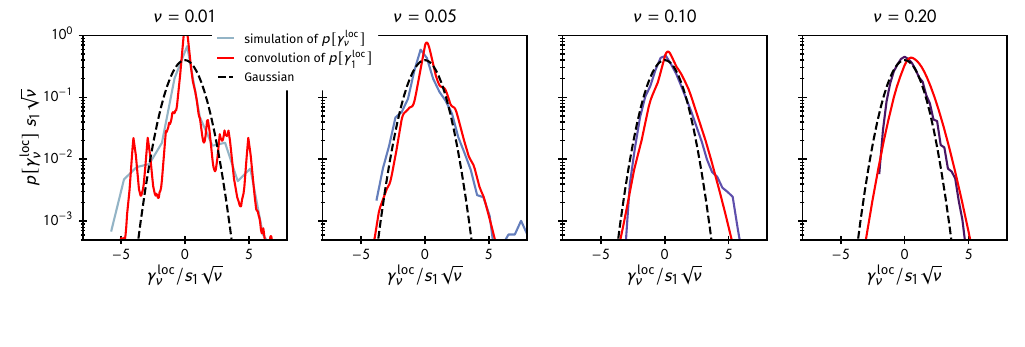}
    \caption{\textbf{The probability distribution of the local fracture energy for multiple missing bonds $\PDF{\dGlocc}$ matches with the convolution of the single-bond probability distributions $\PDF{\dGloc_1}$}
        Probability density functions of normalized local fracture energies $\dGloc$ obtained from direct numerical simulations (blue) and from independent superposition of single bond perturbations (red lines) for different fractions of missing bonds $\fraction$. We use a logarithmic scale for the y axis to better show the tails of the distributions.}
    \label{fig:superposition_distributions}
\end{figure*}

The predicted distribution is in good agreement with the distribution obtained from direct numerical simulations of networks with randomly placed missing bonds (Fig.~\ref{fig:superposition_distributions}), except for a small systematic shift. We attribute this shift to the fact that interactions between removed bonds are neglected when superposing their individual perturbations.

\section{Fit of an exponential to the tail of the $\dGlocc$ distribution}
\label{app:exp-tail-fit}

\begin{figure}
    \centering
    \includegraphics{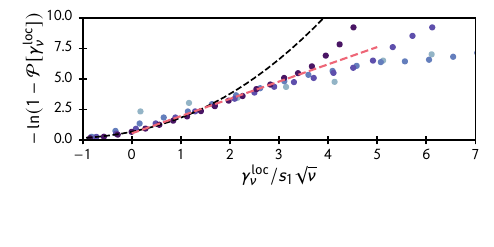}
    \caption{Fit of an exponential to the tail of the cumulative probability distribution of $\dGlocc$. We plotted $\log(1 - \Pcum_{\dGlocc})$ versus $\dGlocc$ so that an exponential distribution appears as a straight line. The dashed red line is the exponential distribution $\Pcum_\mathrm{Exp}$ that we fitted to the numerical simulations (blue dot). The black dashed line corresponds to a Gaussian.}
    \label{fig:sup_exp-fit}
\end{figure}
The exponential cummulative probability distribution is defined as 
\begin{equation}
\label{eq:}
\Pcum_\mathrm{Exp}(\phi) = 1 - e^{- (\phi - \phi_0) / s }~,
\end{equation}
where we fitted the coefficients $\phi_0 = -0.4$ and $s = 0.71$ visually to the data in Fig.~\ref{fig:sup_exp-fit}.

\section{Roughness of the crack path}
\label{app:crack-path}

\begin{figure}
    \includegraphics{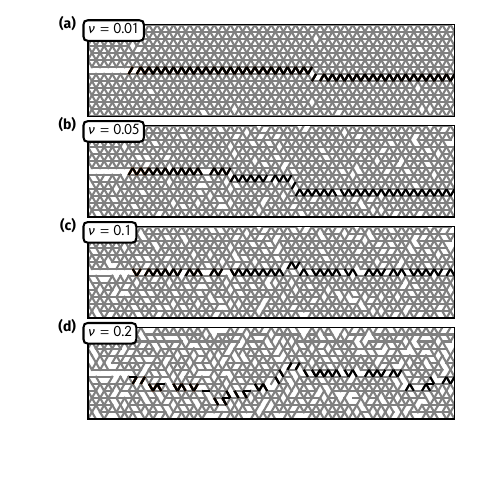}
    \caption{
    \textbf{The crack path is not perfectly straight} 
    Broken bonds (black) on top of the initial network structure (gray)
    for four examples with different concentrations of missing bonds $\fraction$.    
    The network is shown in the undeformed configuration to make deviations from a straight crack path more evident.}
    \label{fig:sup_rough_crack_paths}
\end{figure}

While crack arrest theory presented in the main text assumes a perfectly straight crack path, actual crack paths are rarely perfectly straight in our simulations, see Fig.~\ref{fig:sup_rough_crack_paths}. Even for the smallest fraction $\fraction=0.01$, the crack occasionally jumps to a neighboring row of bonds, and the crack path becomes highly tortuous for $\fraction=0.2$. Even for the roughest case, the crack propagation consists of several straight sections separated by scattering events. Because of the nearly translational invariance of the geom  $\Gloc$ landscape that is statistically similar than in the straight case. 
At the scattering events, the crack deviates from the straight path because an off-the-plane bond breaks at a smaller $G$ than the bond in the crack plane. Hence, the critical energy release rate required to advance the crack is smaller when allowing the crack to deviate from the plane than when forcing it to remain straight~\cite{abid_fracture_2019}.
Therefore we expect our planar crack pinning theory to be an upper bound for the actual fracture energy. However, the numerical results for non-straight crack propagation are virtually indistinguishable from this upper bound, indicating that this effect might be very weak.

\section{Number of broken bonds per unit crack advance}
\label{app:nb-bonds}

\begin{figure}
\includegraphics{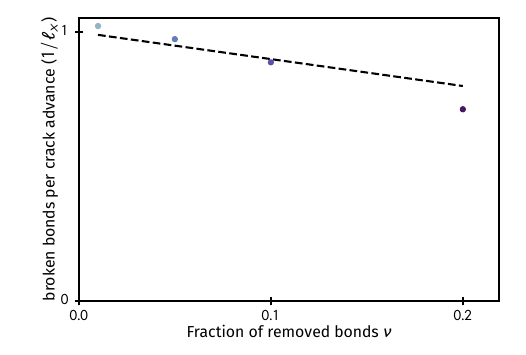}
\caption{\textbf{Number of broken bonds per unit crack advance} as a function of the fraction of missing bonds $\fraction$. 
The number of broken bonds is averaged over a crack propagation length of $800 \lx$ and over 298 realizations. 
The dashed line indicates the expectation for a perfectly straight crack path.}
\label{fig:sup_broken_per_advance}
\end{figure}

It is sometimes argued that crack roughness increases toughness by increasing the number of bonds broken per unit crack advance~\cite{urabe_fracture_2010}.
Our simulations contradict with this expectation: while the fracture energy increases, the number of broken bonds actually decreases with increasing fraction of missing bonds, see Fig.~\ref{fig:sup_broken_per_advance}, despite the increasing crack path roughness.
The decreasing number of broken bonds is expected from the decreased density of bonds in the network, as inidicated by the dashed line.

\end{document}